# Femtoliter Batch Reactors for Nanofluidic Scattering Spectroscopy Analysis of Catalytic Reactions on Single Nanoparticles

*Björn Altenburger[1], Joachim Fritzsche[1] and Christoph Langhammer[1*]*

[1]Department of Physics, Chalmers University of Technology; SE-412 96 Gothenburg, Sweden

*Corresponding author: clangham@chalmers.se

**Abstract**

Macroscopic batch reactors are a core concept in chemical synthesis and catalysis due to their ability to ensure high conversion rates of the used reactants. At the nanoscale, such reactors hold promise due to their potential to enable chemistry in confinement under well-controlled mass transport conditions, and as enablers for the characterization of catalytic reactions on tiny active surface areas, such as single nanoparticles. However, their practical implementation and the readout of reaction products if used for the study of catalytic reactions is challenging due to their tiny volume and the requirement of being able to transiently open and close such nanoscopic batch reactors. Here, we introduce a liquid phase nanofluidic batch reactor concept with a volume of 4.8 femtoliters that conveniently can be opened and closed using a bypassing $N_2$ gas stream, and that in combination with nanofluidic scattering spectroscopy readout enables the characterization of a catalytic reaction on a single nanoparticle inside the reactor, as we demonstrate on the catalytic reduction of fluorescein by sodium borohydride on a Au catalyst.

**Keywords**

**Introduction**

Chemical reactors are the heart of chemical synthesis and catalysis, and of the corresponding industries. They come in many different implementations and designs, with so-called flow and batch reactors probably being the broadest definitions of the two most widely used and principally different reactor principles. In a plug-flow reactor a constant stream of reactants into the reactor and an equally constant flow of product out of it is established. This means that it operates in a continuous steady state, i.e., that reactant feed, temperature and flow rates are all constant to ensure continuous product output. In a batch reactor on the other hand, a chemical or catalytic reaction is conducted in a non-continuous way, such that reactants, products and solvent do not flow in or out of the reactor until the desired conversion has been reached, and the reactor is first emptied and subsequently refilled with fresh reactants – and maybe catalyst – to begin a new reaction cycle.

In their traditional implementations, such reactors are large macroscopic objects operated in chemical plants. However, during the past decades with the advent of microfluidic systems that can control and manipulate fluids in geometrically constrained volumes at the micrometer scale, so-called microfluidic reactors of both the plug-flow and batch type, as well as more complex variants, have been developed. Since their first introduction they have found wide application at both the research and industrial production level due to their intrinsic advantage of high surface-area-to-volume ratios, superior mass and heat transfer control, reduced chemical consumption, and safety[1–9]. Accordingly, today microfluidic reactors find application in disciplines that span from organic synthesis to the production of functional materials and nanoparticles, to enzymatic reactions in the life science domain.

As a natural evolution, with the advent of nanotechnologies enabling the engineering of materials and structures at the nanoscale, chemical reactors have been further downscaled to the regime of so-called nanoreactors, that is, reactors with sub-1 μm dimensions and volumes in the sub-microliter range[10–15]. They enable chemical reactions to occur in nanoconfined space and isolated from the environment. In a wide range of different implementations, they are today applied in (bio)nanotechnology to, e.g., steer chemical transformations and reactions, as well as synthesize complex nanoparticles.

As a simultaneous development, the field of nanofluidics has evolved rapidly and harnesses nanofluidic systems nanofabricated into solid state surfaces to control the flow of fluids at the nanoscale.[16,17] They are today used in widely different areas that include the study and manipulation of biomolecules,[18] osmotic energy conversion,[19] material science,[20] and single cell analysis[21]. Furthermore, it is becoming increasingly clear that the solid-state nature of nanofluidic structures not only enables the control of fluids but offers the additional potential of harnessing, e.g., the electronic, surface chemical or optical properties of the used matrix materials to derive new ways of probing nanofluidic systems.[22–25] At the same time, it is a great challenge in the field to transiently or permanently seal fractions of such a system, such as individual nanochannels, to thereby transiently or permanently isolate the content from the environment, while initially being able to access the same volume conveniently in a flow-through manner to (re)fill it when desired. To address this issue for liquids inside nanofluidic channels, intricate solutions, such as the use of flexible glass[26], deformation[27], Laplace nanovalves[28], graphene seals[29] and thermal bubbles[30] have been reported, and they all have in common that they require both complex nanofabrication and operation infrastructure.

In this work, we introduce nanofluidic batch chemical reactors with femtoliter volumes that neither require additional fabrication nor infrastructure for operation, and that are conveniently combined with nanofluidic scattering spectroscopy (NSS) readout[31] in the visible spectral range. We characterize the operation principle of these reactors using brilliant blue (BB) dye and NSS, and demonstrate their application in single nanoparticle catalysis[32,33], which aims at investigating structure-activity correlations beyond the ensemble average, on the example of the catalytic reduction of fluresceine on a Au catalyst using $NaBH_4$. Specifically, we harness the ability of micro and nanofluidic systems to control gaseous fluids at the nanoscale, as we recently introduced[34–36], while simultaneously also operating the fluidic system in the liquid phase, in which the catalytic reaction occurs on the surface of a Au nanoparticle localized inside a reactant solution filled nanofluidic channel that we transiently isolate into a batch reactor state by establishing a gas flow by either of its ends. Using the catalytic reduction of fluorescein on a Au nanoparticle in $NaBH_4$ aqueous solution as model reaction[37,38], we demonstrate that NSS is able spectrally resolve the intricate dynamics of (de-)protonation and reduction-induced shifts in the light absorption bands of the fluorescein molecule in real time.

## Results and Discussion

*Nanofluidic scattering spectroscopy*

The spectroscopic recording and analysis of light scattered from nanosized objects has developed into a budding field of science as it enables fascinating insights into the world of nanoparticles, molecules and their interactions[39]. To this end, we have recently introduced NSS, a technique that spectrally resolves the light scattered from individual nanofluidic channels and that can be applied to measure the spectral fingerprint of solutes, such as their wavelength-dependent molar extinction coefficient and their concentration, in femto- to attoliter volumes inside a single nanochannel[31]. In this work, we further build on this overall concept and develop a nanofluidic batch reactor by transiently isolating a reactant solution filled nanofluidic channel by establishing a gas flow by either of its ends and combine it with NSS readout to analyze the time evolution of the spectral fingerprint of the catalytic reaction of fluorescein with $NaBH_4$ and a single Au nanoparticle inside the nanochannel.

*Nanofluidic system design*

The fluidic chip used in this work was micro- and nanofabricated into a thermally oxidized Si wafer and the fluidic systems on it were hermetically sealed by bonding an optically transparent glass lid onto the nanostructured oxidized side of the wafer, as described in detail in the **Methods** section. The layout of the fluidic system used is based on our previous work[31] and features a set of parallel sample nanochannels in the center and a corresponding set of colinear reference channels with their own microfluidic inlet system that is not connected to the one of the sample channels. These reference channels are designed to enable online optical referencing in analogy to macroscopic double-beam spectrophotometry, to enable a stable baseline in the NSS measurements (**Figure 1**). Both types of nanochannels are 120 μm long, 200 nm deep and 200 nm wide. To enable single particle catalysis experiments, we nanofabricated Au nanoparticles of 20 nm height, 40 nm width and 640 nm length into the sample channel set used in this work (**Figure 1 a-d**). In addition, each sample channel set also features an empty nanochannel with the same dimensions as the sample channels to enable control experiments without catalyst. On the in- and outlet side, the sample nanochannels are connected to two microfluidic channels with 5 µm height and width to enable efficient fluid exchange (**Figure 1 a-d**). The reference channels, in contrast, are only connected to a single – and independent – microfluidic system on the inlet side, i.e., they are of the dead-end type and filled with the solvent used in the experiment, which in the present work is water (**Figure 1 a-d**). To enable control of the fluid flow through the fluidic system, the fluidic chip as a whole is mounted in a dedicated holder on which Luer-Lock connectors allow the pressurization of the liquid in the inlet reservoirs of the chip using $N_2$ gas or also to establish an $N_2$ gas flow through the fluidic system of the chip if it is dried out, as we also have demonstrated for entirely gas-phase operated nanofluidics[34,35] (**Figure 1e,f**).

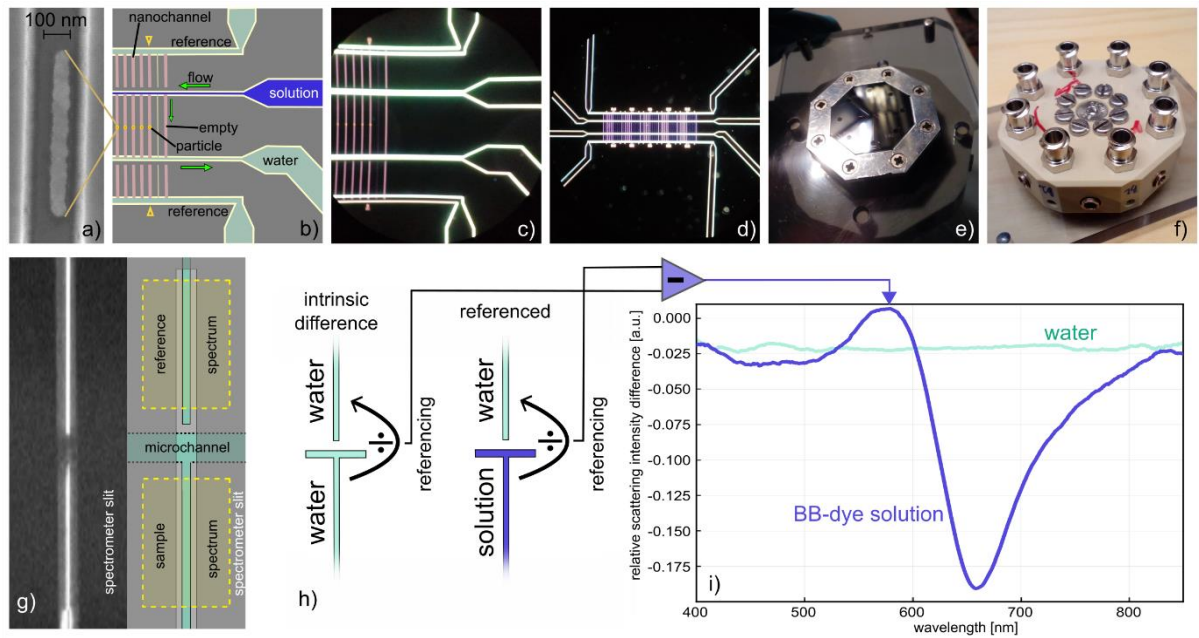

**Figure 1. Fluidic layout and optical referencing scheme for NSS.** *a) Scanning electron microscopy (SEM) image of a representative elongated Au nanoparticle fabricated into the nanochannels. b) Schematic of the micro- and nanofluidic system on the chip. The sample nanochannels are connected to in- and outlet microchannels that enable the exchange of fluid. The reference nanochannels have dead-ends and are colinear with the nanochannels in the center. We note that the sample- and reference fluidic systems are not connected and that we keep the reference system always filled with water during our experiments. The nanochannels on both sample and reference fluidic systems are 200 nm wide and*

*200 nm deep. The sample fluidic system features five channels each with a single 640 nm x 40 nm x20 nm Au particle nanofabricated into their center at the positions indicated by the yellow dots. The nanochannel furthest to the right in the sample system is kept empty. c) Dark-field scattering microscopy image of the fluidic system explained in b) when filled with water. The distance between the nanochannels is 20 μm. The sample channels are 120 μm long and the reference nanochannels are 65μm long. d) A lower magnification dark-field scattering image of the fluidic chip revealing multiple sets of nanochannels. In this work, we used the one discussed in a)-c) only. e) Photograph of the transparent (glass lid) side of the octagonal fluidic chip installed in its holder and used for NSS readout. Some microchannels are visible and connect to the inlet reservoirs on the backside of the chip, which are used to fill the fluidic system with the liquid or gas of choice. f) Photograph of the backside of the holder with circularly arranged screws that enclose access to the reservoirs of the chip and with the Luer-Lock connectors used for pressurizing the fluidic systems with $N_2$ gas to establish convective flow. g) Schematic and dark-field scattering image of a sample and reference nanochannel aligned with the slit of the spectrometer used for NSS readout. The nanochannels are arranged such that spectra from sample and reference channels can be recorded simultaneously, as indicated by the areas marked with yellow. We note that the microchannel connecting to the sample nanochannel is not visible in the dark field image due to the illumination being parallel to the microchannel walls. h) Schematics of the NSS spectra acquisition and referencing scheme discussed in detail in the main text. First, the intrinsic difference between sample and respective reference nanochannels is determined before the solution in the sample channel is exchanged. As a result, the relative scattering intensity difference (RSID) is recorded per wavelength, resulting in so-called RSID spectra. i) RSID spectra obtained for the sample nanochannel filled with water and for 25 mM Brilliant Blue (BB) dye solution, which exhibits the characteristic peaks corresponding to the absorption bands of the dye[31] (see also **Figure S1c**).*

To perform NSS measurements and record scattering spectra from a nanochannel, we align the nanochannel of interest and its closest reference channel within the slit of the spectrometer that is connected to the dark-field microscope (**Figure 1g** – see Methods for details). In this way, we can implement continuous online optical referencing to reduce noise and drift induced by fluctuating light intensities, change of focus of the microscope over time or thermal (expansion) induced effects. The NSS spectrum acquisition and signal treatment sequence is then comprised of the following steps (**Figure 1 g, h**).

(i)  Reference and sample channel systems are filled with the solvent of choice (here: water) and a scattering spectrum is recorded from the sample and reference channel by binning the signal from 21 pixels along the respective nanochannel to reduce noise, which corresponds to a ca. 15 µm long fraction of the entire channel with a volume of ~ 0.6 femtoliter.

(ii)  The obtained water-filled sample channel spectrum is divided by the water-filled reference channel spectrum to obtain the "*intrinsic difference spectrum*" between sample and reference channel. This step is necessary since even two nominally identical nanochannels may exhibit slightly different scattering spectra due to, e.g., slight variations in dimensions or wall roughness.

(iii)  The water in the sample channel system is exchanged by an aqueous solution of the compound of interest (here: BB-dye), by exchanging the liquid in the corresponding reservoir and establishing a flow through the sample nanochannel.

(iv)  The scattering spectrum measured from the sample channel once filled with the dye solution is divided by the simultaneously obtained spectrum of the water filled reference channel.

(v)  The intrinsic difference spectrum obtained in step (ii) is subtracted from the referenced sample spectrum obtained in step (iv) to yield the *relative scattering intensity difference* (RSID) spectrum (**Figure 1i** for a BB-dye example). Such RSID spectra are unique fingerprints of the specific compounds and their concentration in the nanochannel and can be back-calculated to molar extinction coefficient, $\varepsilon(\lambda)$, spectra using the formalism we have introduced earlier[31].

*Conceptual development of a nanofluidic batch reactor*

A key challenge in single particle catalysis experiments, or in any measurement of chemical/catalytic activity where the reaction rate is low or the active surface area very small, is to ensure the accumulation of enough product molecules such that they can be detected, and both identified chemically and quantified per unit time to ultimately enable the derivation of catalytic activity, e.g., as turnover frequency, and selectivity. To this end, as we have demonstrated previously, provided the reaction rate is high enough, nanofluidic systems hosting single catalyst nanoparticles in the tens to hundred nm size range enable such analysis both in the liquid[37,38,40] and gas phase[36,41] in a continuous flow-through fashion where the nanofluidic channel is operated as a plug-flow reactor. At the same time, from these previous works it becomes quite clear that the ability to accumulate product molecules over time in an *enclosed* volume, i.e., a batch reactor, has the potential to significantly increase performance of nanofluidic reactors since it would enable the study of slower reactions, or of even smaller catalyst nanoparticles to ultimately approach the regime of industrial catalysts where few nm particle sizes are very common. Considering the above, it is thus interesting to conceptually develop a transient nanofluidic batch reactor for single particle catalysis and tailored for NSS readout, which we subsequently will implement in practice.

To do so, we first consider the established *plug-flow reactor* scenario (**Figure 2a**), where a catalyst nanoparticle is placed in the center of a nanochannel and the reactant solution is flushed through the channel towards the particle, where it (partly) reacts to a product which then is swept away by the constant convective flow through the channel towards the outlet. This mode of operation means that the actual reactor volume on both sides is connected to large reservoirs with either a high (inlet) or a low (outlet – if not all reactant has been converted by the particle) concentration of reactants. Consequently, reactant molecules can always diffuse in and out of the reactor, with or against the applied convective flow, with the set flow rate dictating the severeness of the effect. This interplay between diffusion and convective flow thus creates intricate reactant/product concentration dynamics inside the nanochannel that are hard to control. Unfortunately, the high convective flows necessary to suppress (back) diffusion of reactants and solvent also lead to excessive dilution of the product and shorter reactant residence times (and thus a smaller fraction of reactants reacting per unit time), which in turn makes it increasingly hard to quantitatively analyze the product formed on the particle. Consequently, it becomes clear that a plug-flow nanoreactor is best suited for experiments in high convective flows over highly reactive particles[42,43].

As the second scenario, we consider a *modified plug-flow reactor* scenario (**Figure 2b**) in which back diffusion of solvent, reactants and product is eliminated by rapidly flushing away the exiting solution at the outlet of the nanochannel in a stream of $N_2$ gas (rather than a stream of pure solvent as in the first scenario). Since, in this scenario, the convective flow of reactant solution through the nanoreactor still is maintained, while back-diffusion is prohibited in principle, sizable accumulation of product necessary for detection in cases of slow reaction rates or tiny catalyst surface area remains challenging, even if convective flow rates can be reduced compared to the first scenario.

The third scenario of the *transient batch reactor* resolves this limitation by not only applying a $N_2$ gas flow by the outlet of the nanoreactor but also by its inlet, once it has been filled with reactant solution (**Figure 2c**). In this way, the enclosed volume contains a constant number of reactant molecules which can react over long timescales on the nanoparticle without risk for dilution by diffusion or convective flow. This in turn means that eventually a high enough product concentration can be obtained such that it can be measured and quantified. It also means that, in principle, the catalyst eventually may convert the entire volume of reactants, provided that the product is stable over time and/or does not poison the catalyst surface. Finally, we note that the system easily can be reopened and refilled by replacing the $N_2$ flow on the inlet side with reactant solution.

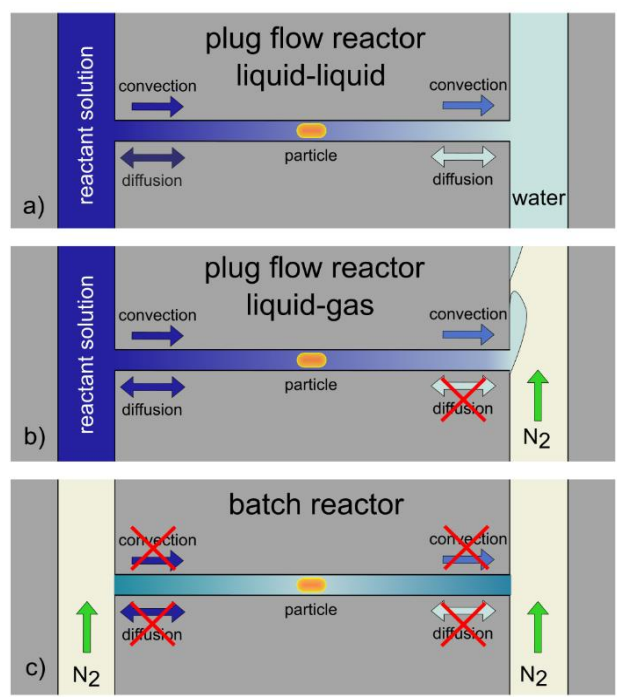

**Figure 2. Nanofluidic reactor concepts for single particle catalysis.** *a) In the liquid-liquid plug-flow implementation of a nanofluidic reactor, a liquid flow along both sides of and through the nanochannel is established. The catalyst particle is positioned in the center of the reactor where it is exposed to a continuous convective inflow of reactants in the solvent of choice (here: water) and a continuous outflow of reaction product and unconverted reactants. In addition to the enforced convective flow, diffusion is*

*also active as the result of the concentration gradients induced by the catalytic turnover of reactants. Notably, since water is continuously flushed through the microfluidic system on the outlet side to carry away reaction product, back-diffusion of water into the nanochannel and against the convective flow through it also takes place, thereby inducing an uneven product concentration along the nanochannel. b) In the liquid-gas plug-flow implementation, the microfluidic system on the outlet side of the nanochannel is connected to a gas (here: $N_2$) to replace water (the solvent) as the medium to carry away the product-reactant-solvent mixture exiting the nanoreactor. This has the advantage that back-diffusion of solvent is suppressed. c) In the batch reactor implementation in focus of this work, an $N_2$ gas flow is established through the microfluidic systems on both the inlet and outlet sides of the nanochannel, once it has been filled with reactant solution. In this way, not only solvent back-diffusion is suppressed, but the entire reactor volume is transiently sealed off.*

As the next step, we experimentally verify the above conceptual discussion of the different reactor types by analyzing the spatial distribution of the concentration of a BB-dye solution along an empty nanochannel (i.e., no catalytic particle yet) in the first two scenarios. For the liquid-liquid plug-flow reactor scenario (cf. **Figure 2a**) we establish a continuous flow of BB-dye solution through the empty nanochannel by applying a relatively low 100 mbar pressure difference across it when the inlet fluidic system is filled with a 25 mM BB-dye solution and the outlet fluidic system with water (**Figure 3a**). We then define six sections along the nanochannel (labelled 1 to 6 in **Figure 3a**) from which we acquire NSS RSID spectra as outlined above, by binning 21 pixels for each section. Clearly, while all six RSID spectra show the distinct characteristics of the BB-dye[31] it also is obvious that the spectral features exhibit different intensities that systematically depend on the position along the channel (**Figure 3b**). Intensity is highest closest to the inlet (section 1) and lowest closest to the outlet, i.e., the water filled fluidic system (section 6), which indicates a distinct concentration gradient in the direction of convective flow. This becomes even more evident when plotting the time evolution of the amplitude of the strong negative peak at 657 nm, which relates to the strong absorption band of the BB-dye (**Figure S1c**), after starting the inflow of dye at t = 0 s (**Figure 3c**). It reveals that all monitored channel sections respond to the inflow of the dye into the nanochannel at t = 300 s, but that the sections further away from the inlet react more sluggishly and never reach the same RSID peak amplitude. This means that the nominal BB dye concentration is never reached in the downstream channel sections due to back diffusion of water from the outlet system and the corresponding continuous dye dilution in the nanochannel.

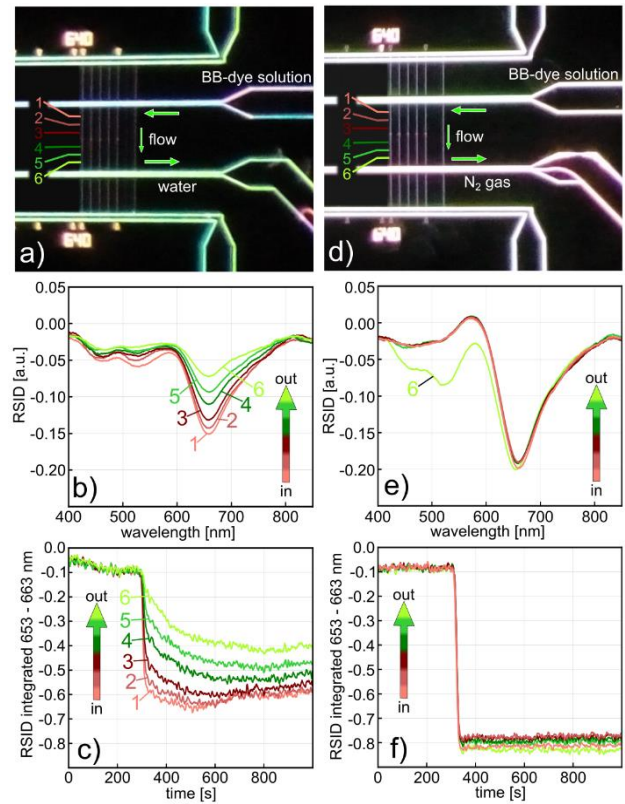

***Figure 3. Liquid-liquid and liquid-gas plug flow reactors.*** *a) Liquid-liquid plug-flow nanofluidic reactor. Dark-field scattering microscope image of an entirely liquid-filled fluidic system, where the inlet microfluidic system and the sample nanochannels are filled with 25 mM BB-dye solution and everything else with water, i.e. a convective water flow through the microfluidic outlet system is established to flush away BB-dye solution exiting the sample nanochannels. For the NSS readout, the empty (i.e., no particle) sample nanochannel used here is divided into six 21 pixel sections (0.6 femtoliter volume each) evenly distributed along the channel, from all of which NSS spectra are recorded simultaneously using the referencing scheme introduced in* **Figure 1h**. *b) Applying pressure to the inlet and microfluidic system filled with BB-dye establishes convective flow of the dye through the nanochannel, as corroborated by the distinct BB-dye fingerprint of the RSID-spectra measured along the channel using NSS. Notably, the spectra measured from nanochannel sections closer to the water-filled outlet side show a lower intensity of the negative RSID peak at 657 nm, as a consequence of water back-diffusion (and thus BB-dye dilution) form the water-filled outlet microchannel. c) Time-traces of the negative 657 nm negative RSID peak amplitude (calculated by integrating 10 spectral datapoints around the peak) for the six sections of the nanochannel. The trace from section 1, which is closest to the BB-dye solution filled inlet microchannel, shows a relatively rapid decrease when the dye solution flow is initiated at 300 s. In contrast, the trace from section 6, which closest to the water/outlet*

*side, shows a delayed and weaker RSID-peak response that indeed corroborates significant dilution of the BB-dye solution by water that enters the nanochannel via back-diffusion against the convective flow, since RSID-peak intensity is proportional to BB-dye concentration[31]. d) Liquid-liquid plug-flow nanofluidic reactor. Dark-field scattering microscope image of the fluidic system of the chip, where the reference fluidic system is filled with water, the inlet microfluidic system and the sample nanochannels are filled with 25 mM BB-dye solution and a convective $N_2$ gas flow through the microfluidic outlet system is established to flush away BB-dye solution exiting the sample nanochannels. e) Same as b) but for $N_2$ flow on the outlet side. Clearly, the RSID spectra of the six channel sections are now very similar, indicating similar BB-dye concentrations along the entire nanochannel since water back-diffusion is efficiently suppressed. The reason for the different appearance of the RSID-spectrum from section 6 is that the closely adjacent $N_2$-filled microchannel scatters more light than when it is water filled, meaning the optical referencing is not perfect. f) Same as c) but for $N_2$ flow on the outlet side. Notably, all six time traces now drop instantaneously and simultaneously as the BB-dye enters the nanochannel at 300 s, and all traces reach the same RSID-amplitude level, which indicates identical dye concentration along the entire nanochannel, which also stays constant over time.*

To address this issue, we implement the second scenario (cf. **Figure 2b**), which means that instead of a continuous water flow through the outlet fluidic system, we establish a flow of $N_2$ gas by the outlet of the nanochannel by applying a pressure of 2000 mbar $N_2$ to an empty reservoir at the outlet side of the sample fluidic system (**Figure 3d,** also **Figure S2b**). As evident from the RSID spectra of the six sections along the nanochannel, all peaks now align at the same amplitude, corroborating that the BB dye concentration is identical along the whole channel and that back-diffusion, and thus dilution by water from the outlet microchannel, is suppressed (**Figure 3e**). Only the spectrum from section 6 closest to the $N_2$-filled microchannel appears different in the short wavelength range. This is the consequence of the significantly increased light scattering intensity of the $N_2$-filled microchannel compared to the water-filled one, and the consequent imperfect optical referencing. The suppression of back-diffusion of water is further corroborated by again plotting the time evolution of the amplitude of the negative RSID peak at 657 nm after starting the inflow of BB-dye at t = 0 s (**Figure 3f**), in which now all six sections perfectly overlap and remain constant during the entire measurement.

*Implementing a nanofluidic batch reactor*

Motivated by the results summarized in **Figure 3**, we move forward to implement and analyze the third scenario, that is, a transient batch reactor enabled by enclosing a liquid-filled nanochannel between two $N_2$ gas streams by flushing $N_2$ through the in- and outlet microchannels that connect to the nanochannels (**Figure 4**). Implementing this function combined with NSS readout necessitates the following subsequent steps (see also **Figure S2**).

(i)   All micro- and nanofluidic channels in the sample fluidic system are flushed with $N_2$ gas that is provided via empty inlet reservoirs on one side of the microchannels that connect to the nanochannel on both its ends (**Figure 4a**). At the same time, the nanofluidic system used for referencing is filled with water.

(ii)  To measure the intrinsic difference spectra between sample and reference channels necessary for NSS, the sample fluidic system is filled with water by applying pressure on the water-filled reservoir connected to one end of the microchannel on the chip outlet side using a syringe (**Figure 4b**). Importantly, in this way if the pressure applied to the water reservoir is higher than the applied $N_2$ pressure, water is pushed into the system until it reaches a nanochannel that then also fills with water. Elegantly, the spatial extension of water-filling in the system can be finely tuned by balancing the pressures applied to the $N_2$ side and the water side. Hence, because in any nanochannel that is water-filled in this process the overall pressure on the water side is higher than in the outlet side that connects to a microchannel through which an $N_2$ flow still is established, the water will flow through the nanochannels. The intrinsic optical difference spectra between sample and reference channel can therefore now be measured.

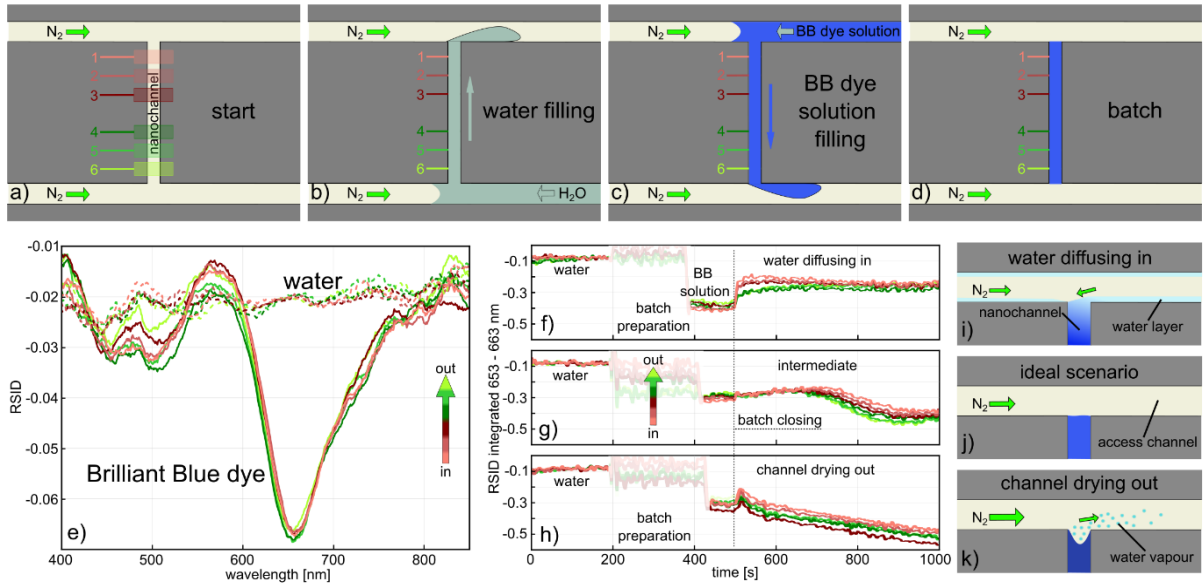

*Figure 4. Creating a nanofluidic batch reactor with NSS readout.* Schematic depiction of the key steps, for which an overview over the whole chip is presented in **Figure S2**. *a) At start, the system is entirely flushed with $N_2$ gas. Six different channel sections are used for NSS readout. b) Applying pressure on the water-filled inlet reservoir that is higher than the $N_2$ pressure applied to the second inlet reservoir on the opposite side of the lower microchannel establishes a convective water flow into the microchannel which, once it reaches a nanochannel, also leads to water flow through that nanochannel. The water exiting the nanochannel is carried away by the $N_2$ stream in the upper microchannel. In this state, the intrinsic difference in light scattered from the water filled sample nanochannel and its reference channel can be determined as the first step of the NSS measurement process. c) After flushing out all the water from the system with $N_2$ gas by lowering the pressure applied to the water-filled reservoir, the process described in b) is now repeated but from the opposite direction and using a BB-dye solution filled reservoir until the nanochannel is filled with BB-dye. d) By re-establishing the $N_2$ gas flow through both microchannels, the BB-dye solution is contained inside the nanochannel and the batch reactor is "closed". e) NSS-RSID spectra measured at the six sections of a nanochannel depicted in a) for a water-filled (condition depicted in (b)) and BB-dye-filled (condition depicted in (c)) nanochannel. f-h) Time traces of the BB-dye negative RSID-peak area integrated between 653-663 nm for all channel sections and for three different $N_2$ flow scenarios achieved by applying 1380 mbar, 1450 mbar, 1600 mbar of $N_2$ pressure, respectively. Until t = 200 s, the system is water-filled. Between 200 s and 460 s, the batch reactor is filled-up with BB-dye solution and the light path to the camera is blocked (shaded area). After 460 s, the system is completely filled with BB-dye solution and closed-off at 500 s. i-k) Schematics explaining the integrated RISD-peak area time traces in (f-h). i) At low $N_2$ flow rates the mayor volume of the outlet microchannel is depleted of water while a thin water layer remains on the channel walls due to the hydrophilicity of $SiO_2$. The diffusion of water from this layer into the nanochannel explains the decrease in negative RISD amplitude in (f). k) At the highest $N_2$ flow rate,*

*water is rapidly evaporating and to the extent that the nanochannel starts to dry out, thereby increasing the BB-dye concentration inside it. This explains the increase in negative RISD amplitude in (h). j) In an ideal situation, the two effects above are in equilibrium for an extended period of time before the drying out of the nanochannel sets in. This is the case in the negative RISD amplitude time trace in (f) for 1450 mbar applied $N_2$ pressure.*

(iii) After the intrinsic difference spectra collection, the water is removed from the sample fluidic system by releasing the pressure on the syringe and thus establishing $N_2$ gas flow through the entire fluidic system once again. Subsequently, the sample nanochannels are filled with 25 mM BB-dye solution. To do this, we repeat the same process as in step (ii) but from the opposite side by this time applying pressure higher than the $N_2$ one onto a dye solution filled reservoir (**Figure 4c**). In this way, the nanochannel can be filled with dye solution in a controlled fashion by establishing a net-flow. This situation corresponds to the second scenario discussed above (cf. **Figure 2b**)

(iv) As the final step to establish a batch reactor condition the overpressure on the dye solution reservoir is released to re-establish the $N_2$ flow through the microchannel on the inlet side (**Figure 4d**). In this way, the $N_2$ flow will push the dye solution back out of the microchannel, while the nanochannel remains filled with BB-dye solution since the same $N_2$ pressure is applied on both sides of the fluidic system and thus across the nanochannel.

Having conceptually established this procedure, it is now interesting to monitor it step-by-step using the NSS readout from the six sections along the nanochannel. Accordingly, **Figure 4e** shows NSS-RSID spectra measured for a sample nanochannel filled with water (c.f. scenario in **Figure 4b**) and 25 mM BB-dye solution (cf. scenario in **Figure 4c**). As expected, the RSID spectra are very similar in all six sections and correspond either to a flat line (water) or exhibit the strong negative RSID peak at 657 nm that signifies the strong absorption band of the BB-dye.

To analyze the batch reactor formation over time, we track the main negative RSID peak amplitude integrated from 653 nm - 663 nm for the six nanochannel sections in three slightly differently executed experiments in which we in the final step varied the applied $N_2$ gas pressure (1380 mbar, 1450 mbar, 1600 mbar, respectively) and thus the flow through the microchannels by both exits of the nanochannel (**Figure 4f-h**). In the first stage, from t = 0 s to t = 199 s, when the nanochannel is filled with water (step (II) described above), the time trace of the RSID peak amplitude stays fairly constant and is at the same level for all six nanochannel sections and in all three experiments, as expected (**Figure 4f-h**).

Between t = 200 s and 440 s, the water in the nanochannels is exchanged with the BB-dye solution by manually applying finely tuned pressure on the BB-dye solution reservoir with a syringe according to step (III) described above. This step requires the direct observation of the fluidic system through the eyepiece of the microscope. Hence, during this period, the acquired RSID data are unreliable and thus not analyzed (shaded area in **Figure 4f-h**). At t =441 s, the nanochannel is completely filled with the BB-dye solution that continuously flows through it, i.e., step (III) is completed, and the RSID-trace is reliable again (shaded area ends in **Figure 4f-h**).

At t = 500 s, we close the batch reactor using the procedure described in step (iv). We observe distinctly different RSID traces for the three different experiments with the three different applied $N_2$ pressures (**Figure 4f-h**). For the lowest applied $N_2$ pressure (1380 mbar), the negative amplitude of the RSID peak at 657 nm, which is proportional to the BB-dye concentration,[31] decreases almost immediately upon batch reactor closing in all six nanochannel sections (**Figure 4f**). Subsequently, it stays relatively constant until approx. 600 s, when a slow increase sets in. This response can be interpreted as follows. While the established $N_2$ flow along the nanochannel in- and outlet indeed efficiently and immediately cuts off the convective flow of BB-dye solution through the nanochannel, a thin water layer remains at the channel walls for some time due to the hydrophilicity of $SiO_2$ (**Figure 4i**). Until this layer is completely dried-off by the $N_2$ flow, a low level of water back-diffusion into the nanochannel is enabled and the reason for the observed decrease of integrated negative RSID-peak amplitude.

To test this hypothesis, we look at the measurement where we have increased the applied $N_2$ pressure to 1450 mbar and thus establish a significantly higher $N_2$ flow through the microchannels (**Figure 4g**), which we expect to dry off any remaining water more rapidly (**Figure 4j**). Indeed, we see that the initial decrease of the 657 nm RSID peak amplitude is essentially absent and replaced by a very slow and much smaller decrease until t ~ 650 s, beyond which the peak amplitude increases again. This increase indicates a dye concentration increase in the nanochannel. However, since no dye is supplied to the nanochannel in this stage of the experiment, we argue that it is the consequence of water evaporating from the nanochannel via the high $N_2$ flow, thereby locally increasing dye concentration in the channel. This finding suggests that the applied $N_2$ flow rate is a critical parameter when closing the batch reactor. To corroborate this effect, we therefore further increased it by applying 1600 mbar $N_2$ pressure in the last experiment (**Figure 4h**). Evidently, the 657 nm RSID peak amplitude time traces now reveal a monotonously increasing negative amplitude, that indicates an increase of the BB-dye concentration inside the enclosed nanochannel due to rapid evaporation of water at the nanochannel in- and outlet into the high-flow $N_2$ stream (**Figure 4k**).

*Using the batch reactor for the catalytic reduction of fluorescein on an Au nanoparticle catalyst*

We have demonstrated in our previous works that a catalytic reduction reaction of fluorescein takes place on the surface of single nanofabricated[37] or colloidal[38] Au nanoparticles inside plug-flow nanofluidic reactors in the presence of $NaBH_4$ reducing agent. In these works, we relied on the quenching of fluorescent emission in the reduced state of the fluorescein molecule as the readout to assess catalytic activity using fluorescence microscopy. Here, we choose to use the same system to apply the nanofluidic batch reactor concept developed above to a catalytic reaction on a single Au nanoparticle (40 nm width, 640 nm length, 20 nm height, **cf. Figure 1a**), and to investigate it using the spectroscopic NSS readout in a nanochannel with 200 x 200 $nm^2$ cross section and a total length of 120 μm, which corresponds to a total solution volume of 4.8 femtoliters (fl) contained inside it. Using NSS readout means that we are not relying on the fluorescent emission signal, but rather on the light absorption spectral fingerprint of the system[31].

We executed the experiment using a reactant solution of 25 mM fluorescein and 120 mM $NaBH_4$ in water (**Figure 5**), and focused on two channel sections, A and B, on either side of the nanoparticle (**Figure 5a**) for the NSS analysis, by again binning 21 pixels for each section.

For this purpose, we performed the batch reactor formation according to the procedure detailed in **Figure 4b-d** and executed a measurement sequence in the same way as outlined above for the BB-dye solution (**cf. Figure 4h,i**). As the first analysis step, we traced the RSID intensity integrated between 518 and 528 nm over time (**Figure 5a**), i.e., around the minimum of the strong negative RSID peak of fluoresceine that corresponds to its main absorption band[31,44] (**Figure S2d**). The first 200 s correspond to the flushing of water through the nanochannel and reveal identical integrated RSID amplitude baselines for both sections (**Figure 5a**). Subsequently, from 200 to ca. 450 s, we exchanged the water with the reactant solution, which again required monitoring of the sample through the eyepiece (shaded area in **Figure 5a**). Subsequently the system was stabilized in the state where a constant convective flow of reactants through the nanochannel is established, and where a $N_2$ gas flow carries away the exiting reactants on the outlet side. Notably, both channel sections deliver identical RSID response. At t = 520 s, we closed the batch reactor by establishing $N_2$ gas flow on both sides of it with an applied pressure of 1450 mbar, which gives rise to distinct reduction in negative RSID amplitude over time in both monitored sections.

To understand this behavior, it is interesting to look at a selection of full RSID spectra taken before and after the batch reactor was closed, i.e., from t = 500 s to t = 540 (**Figure 5b**), which corresponds to the system being in a state of continuous convective reactant inflow (t = 500 s) and 20 s after closing of the batch reactor (t= 540 s). At t = 500 s, the RSID spectra look very similar in both channel sections. They are characterized by a distinct negative RSID peak at ~ 520 nm, which signifies the strong absorption band of fluorescein in an alkaline environment[45], as the case here for the used $NaBH_4$ solution (see **Figure S1d** for a comparison of fluorescein RSID-spectra measured in aqueous solution with and without $NaBH_4$). In the spectra taken after the batch reactor was closed until t = 540 s, we see two major new features emerge. Firstly, the main negative RSID peak at ~ 520 nm has decreased in intensity, which indicates a lower concentration of fluorescein species in their initial state in the reactant solution. At the same time at the position of 470 nm, a significant decrease in RSID occurs, and a distinct new negative peak has emerged (**Figure 5b**). This indicates the formation of a chemically different fluorescein species in the solution, with an absorption band at shorter wavelengths.

It is therefore interesting to follow the time evolution of the full spectrum in detail for the entire course of the experiment until t = 1000 s. Since both channel sections respond very similarly, we select one of them, B, as representative for the whole system (see **SI Figure S4** for the corresponding response of section A) and plot the time series of the inverted RSID spectra from 460 s onward at constant time intervals of 10 s (**Figure 5c**). Evidently, after closing of the batch reactor the main peak at ~ 520 nm rapidly diminishes while the new peak at ~ 470 nm emerges almost instantaneously. However, the 520 nm peak never completely vanishes but only decreases in intensity over time to reach a minimum beyond which it slowly increases in intensity again. Furthermore, we notice that the peak at 520 nm appears slightly shifted to shorter wavelengths in the latest spectra compared to the very first one in the displayed time series. In contrast, the 470 nm peak slowly and monotonously decreases over time after its initial fast appearance. As the final observation, we note that the broad RSID feature observed between 570 nm and 800 nm also decreases in intensity over the course of the experiment. Since we can attribute this feature to the presence of non-absorbing species in the water-based reactant solution, such as $NaBH_4$ and its subsequent products in the present case (**Figure S1b**), and since the intensity of this feature is proportional to the concentration of these species[31], its observed decrease over time signifies a decrease in $NaBH_4$ species, i.e., their conversion into a different species.

To further investigate the negative RSID peaks at ~520 nm and ~ 470 nm in detail, we plot their time evolution from t = 400 s in 2 s time steps by again integrating the amplitude 518 - 528 nm (**Figure 5d**) and 465 - 475 nm (**Figure 5e**), respectively. This analysis corroborates the relatively slow and continuous amplitude decrease of the long wavelength RSID peak until ~ 800 s, beyond which it slowly increases again. It also corroborates the rapid growth of the short wavelength RSID peak upon closing of the batch reactor (on the order of 8 s), which is followed by a slow decrease in amplitude until a plateau is reached at ~ 800 s. Furthermore, tracing the spectral position of the long-wavelength peak over time, reveals that it blue-shifts from ~ 520 nm to ~ 510 nm after closing of the batch reactor over a time scale that is very similar to the aforementioned amplitude decrease (**Figure 5f**). At the same time, we find that the short-wavelength peak slightly shifts to longer wavelengths, i.e. from ~ 470 nm to ~ 495 nm. Finally, we also note that repeating an identical experiment using an empty batch reactor without Au nanoparticle as a control, confirms the catalytic processes on the Au nanoparticle surface as the source of the observed time evolution of the spectral features in the NSS spectra (**Figure S3**).

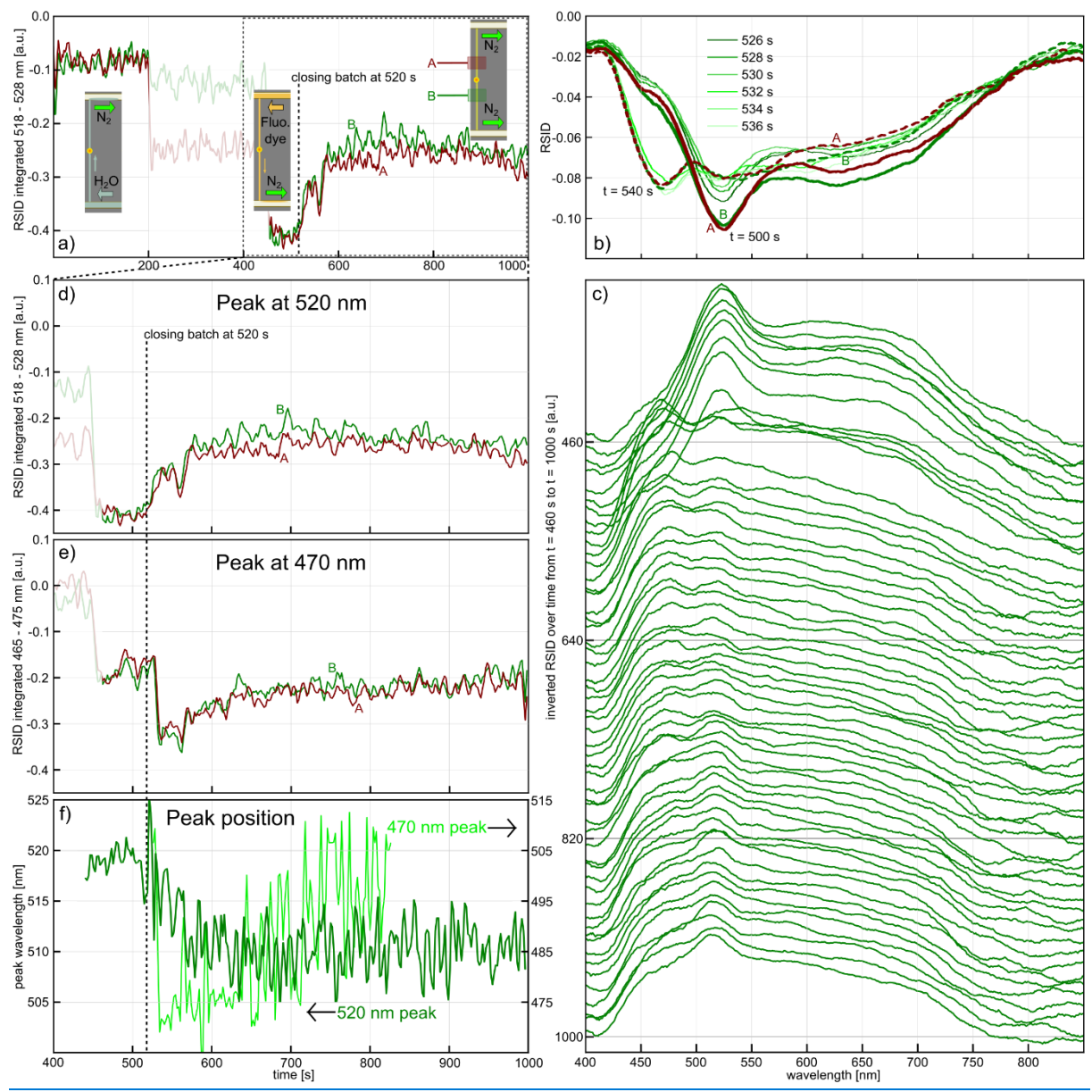

***Figure 5. Fluorescein reduction in a nanofluidic batch reactor over a single gold nanoparticle.*** *a) Time traces of the negative RSID-peak amplitude integrated from 518 – 528 nm for the two monitored sections of the nanochannel, A and B (inset), depicting the flushing of water through the nanochannel (0 – 200 s, **cf. Figure 4b**), the filling of the nanochannel with 25 mM fluorescein and 120 mM NaBH₄ aqueous solution (400 – 520 s, **cf. Figure 4c**), and the operating closed batch reactor (520 – 1000 s, **cf. Figure 4d**). The shaded area depicts again the period when the system is observed through the microscope eyepiece. b) RSID spectra taken before (t = 500 s), at (t= 520 s), and after (until t= 540 s)*

*the batch reactor was closed. At first, they are characterized by a distinct negative RSID peak at ~ 520 nm, which signifies the strong absorption band of fluorescein in an alkaline environment until a second peak starts to emerge at ~ 470 nm upon closure of the batch reactor. This indicates the formation of a chemically different fluorescein species in the solution, with an absorption band at shorter wavelengths. c) Ridge plot of the inverted RSID spectra taken from t = 460 s, when reactant solution is flowing through the nanochannel, to 1000 s, when the batch reactor had been closed and working for 480 s. d) Detailed view of the time trace in (a) from 400 s to 1000 s. It is evident that the integrated negative RSID peak amplitude decreases after the closing of the batch reactor, but also that it increases again slightly after 800 s. e) Same as d) but for the newly emerging peak at ~ 470 nm. In contrast to the ~ 520 nm peak, a rapid rise in amplitude occurs, followed by a steady decline. f) Time evolution of the spectral position of the long- (left y-axis) and short-wavelength (right y-axis) peak maxima.*

*Deriving the reaction mechanism*

Based on the above analysis of the time evolution of the spectral features in the RSID-spectra recorded from the batch reactor, we make an attempt at deriving the reaction mechanisms at play. For this purpose, we first converted the RSID spectra obtained during the experiment into molar extinction coefficient spectra, $\varepsilon(\lambda)$, using the formalism we have introduced earlier[31], and plot a selection of such spectra along the experiment timeline (**Figure 6a**). For comparison, we also include $\varepsilon(\lambda)$ spectra of 25 µM fluoresceine measured in an alkaline NaOH solution (pH = 12.6) and in an acidic HCl solution (pH = 1.3) using a standard double-beam spectrophotometer (Varian Cary 5000) and a 3.5 ml cuvette. Clearly, there is a striking similarity between the $\varepsilon(\lambda)$ spectra obtained by NSS from the nanofluidic batch reactor, and the spectra obtained using standard vis-spectrophotometry, where the early spectra from the NSS measurement agree well with the alkaline, and the late spectra with the acidic environment. This suggests that a significant pH change occurs in the batch reactor during the experiment. To this end, the evolution of our $\varepsilon(\lambda)$ spectra from the batch reactor is in good agreement with Lapierre et al.[45], who have demonstrated that the main absorption peak of fluorescein is localized at ~ 500 nm in highly alkaline environment and that it systematically shifts to ~ 440 nm in strongly acidic environments.

To derive the reaction mechanism at play, we first remind ourselves that the system at hand is comprised of an aqueous solution of fluorescein (25 mM) and $NaBH_4$ (120 mM), and a single Au nanoparticle catalyst. Furthermore, we note that $NaBH_4$ undergoes an auto-decomposition reaction (i.e., without involving the Au nanoparticle) in water, which corresponds to the blue-coded pathway indicated in **Figure 6b** that we have adapted from Gonçalves et al.[46] The first step of this auto-decomposition includes the generation of hydroxide ions ($OH^-$), which cause an increase in the pH of the solution. This thereby increasingly alkaline environment eventually not only inhibits further $NaBH_4$ auto-decomposition but also deprotonates the fluorescein. Hence, it creates the alkaline environment responsible for the strong fluorescein absorption peak at ~500 nm in the $\varepsilon(\lambda)$ spectra (corresponding to the negative RSID peak at ~ 520 nm in **Figure 5b**) measured before closing of the batch reactor at t = 520 s (**Figure 6a**). The subsequent steps of the auto-decomposition reaction are the formation of $B(OH)_3$ (boric acid) and finally of $NaB(OH)_4$ (sodium tetrahydroborate). Notably, this reaction takes place and reaches its equilibrium in the reactant solution already when it is prepared, i.e., before it is flushed into the nanoreactor.

On the surface of the Au catalyst nanoparticle, a second decomposition reaction of $NaBH_4$ can take place via a series of steps that are expected to occur at a high rate (black pathway in **Figure 6b**).[47,48] Most importantly, each of these steps consumes a hydroxide ion, and produces an electron and a proton ($H^+$). We therefore argue that this reaction gives rise to a rapid change of the pH inside the batch reactor back from alkaline to acidic, once the reactor is closed and fresh supply of reactant solution is cut off. As a consequence of this pH change, fluorescein in solution is protonated, which gives rise to the rapid emergence of the new peak at ~ 440 nm that we observe in in the $\varepsilon(\lambda)$ spectra derived from RSID.

At the same time, fluorescein is directly catalytically reduced on the surface of the Au catalyst by the hydrogen radicals that form during the catalytic decomposition of $NaBH_4$ that simultaneously also occurs on the Au.[38] However, the rate of the catalytic decomposition of fluorescein is expected to be significantly lower, due to its low turnover frequency on the order of 0.04 per site and second[37]. This means that two reactions occur at the same time on the Au nanoparticle surface: (i) the rapid consumption of the hydroxide ions available in the solution from the auto-decomposition of $NaBH_4$ via that Au-catalyzed decomposition of (not auto-decomposed) NaBH$_4$ that leads to a rapid change from alkaline to acidic conditions, and (ii) the slower catalytic decomposition of fluorescein using the H-radicals generated in (i).

Having established the different reactions we expect to be at play in the batch reactor initially, it is interesting to discuss their implications for the expected time evolution of the $\varepsilon(\lambda)$ spectra. Starting from an alkaline environment due to the auto-decomposition of NaBH$_4$ in the reactant solution that is the reason for the strong absorption peak at ~ 500 nm prior to (t = 460 s ) and just after (t = 520 s ) closing of the batch reactor (**Figure 6a**), we argue that the rapidly emerging ~ 440 nm peak is the consequence of a rapid change from alkaline to acidic pH due to the consumption of OH$^-$ in the catalytic decomposition of NaBH$_4$ on Au. The subsequently observed slow decrease of the ~ 440 nm peak over time (**Figure 6a**) is then the consequence of the slower catalytic reduction of the protonated fluorescein species on the Au catalyst into a transparent (i.e. non-absorbing) product, as reported by Wang et al[44]. Alternatively, it could be indicative of a slow change of the pH in the reactor towards a more alkaline environment, which, in principle could happen once the hydroxide ion concentration in solution has been significantly lowered and thus (i) the NaBH$_4$ decomposition rate over the Au particle is very low and (ii) the NaBH$_4$ auto-decomposition in water again takes place at a sizable rate.

Finally, the slight and slow increase in intensity of the ~ 500 nm peak in the $\varepsilon(\lambda)$ spectra towards the end of the experiment (**Figure 6a**) may also have two reasons: (i) a change back to higher pH or (ii) re-oxidizing of catalytically reduced fluorescein by oxygen species (red pathway in **Figure 6b**) supplied to the batch reactor via the N$_2$ gas that is flushed by its entrances  (see **Figure S5** for an analysis of the gas composition). As a final note on **Figure 6b** and to connect back to **Figure 5f**, we argue that the final pH of the reactant solution is not as alkaline is it was before encountering the gold nanoparticle, as the residual RSID peak above 500 nm remains shifted to 510 nm and the OH$^-$ producing NaBH$_4$ has been fully consumed by the reaction.

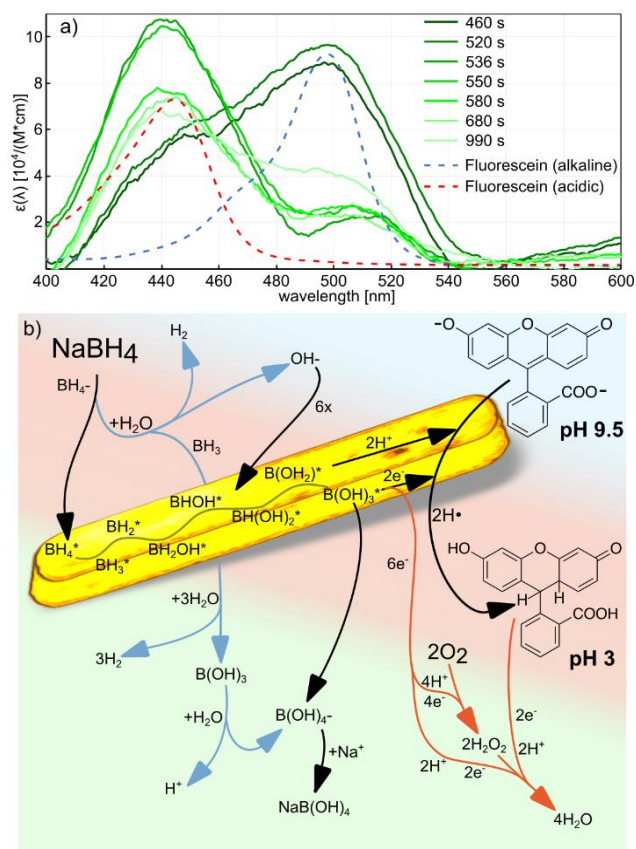

**Figure 6. Molar extinction coefficient, $\varepsilon(\lambda)$-spectra and proposed reaction scheme.** *a) Selected $\varepsilon(\lambda)$-spectra calculated from RSID spectra shown in **Figure 5c** using the formalism we have stablished earlier;[31] as obtained along the timeline of our experiment (green lines). Also shown are fluorescein $\varepsilon(\lambda)$-spectra measured in an aqueous acidic (pH 1.32, HCl, red dashed line) and alkaline (pH 12.54, NaOH, blue dashed line) environment. b) Proposed reaction schemes for the processes occurring in the batch reactor. The blue arrows outline the auto-decomposition pathway of $NaBH_4$ in water, without the influence of a catalytic particle. $NaBH_4$ dissociates and the borohydride ion ($BH_4^-$) reacts with water to form hydrogen ($H_2$), Borane ($BH_3$) and a hydroxide ion ($OH^-$). These hydroxide ions shift the pH to ~ 9.4 (for a 120 mM solution) which induces the deprotonation of fluorescein in solution. Subsequently, sodium tetrahydroxyborate is formed as the final product.[46] The same product is also formed via a second pathway, the catalytic $NaBH_4$ decomposition on the Au catalyst nanoparticle[47,48] (black pathway). However, in this pathway, the decomposition steps consume one hydroxide ion each and produce one electron each, together with a $H^+$ ion on the catalyst surface. These hydrogen radicals can react with an adsorbed fluorescein, thereby reducing it and shifting its absorption band out of the visible spectral range[44]. At the same time, the consumption of hydroxide ions from solution will shift the pH towards acidic. The red pathway depicts the re-oxidation of reduced fluorescein by oxygen species dissolved in the reactant solution from $O_2$ contamination in the $N_2$ gas stream (**Figure S5**) applied to close the batch reactor.*

## Conclusions

We have introduced the concept of nanofluidic batch reactors with a volume of only 4.8 femtoliters that can be transiently opened and closed by combining liquid and gas flows, and how they can be integrated in a silicon-based micro- and nanofluidic system. To probe both the general batch reactor concept and the catalytic reaction of fluorescein with $NaBH_4$ reducing agent on a single Au catalyst nanoparticle inside the batch reactor, we applied nanofluidic scattering spectroscopy, NSS. This allowed us to spectroscopically monitor the presence and exchange of fluids inside the batch reactor in the visible spectral range with high spatial resolution, as well as to follow the catalytic reaction process by monitoring the spectral fingerprint of fluorescein in real time as the reaction evolved. Using this analysis scheme, we found a reaction mechanism that involves multiple pathways in parallel and where the observed changes in the fluorescein molar extinction coefficient over time derived from NSS measurements are caused by (i) significant and dynamic changes in the pH of the reactant solution inside the batch reactor, which shifts the absorption bands of fluorescein, and (ii) the catalytic reduction of fluorescein to a non-light-absorbing species on the Au surface.

In a wider perspective, our results advertise the nanofluidic batch reactor concept, in concert with NSS or potentially other readouts, such as Raman or IR-spectroscopy, for the study of slow catalytic reactions on single nanoparticles. They also hold promise to open the door to single particle catalysis experiments on colloidal[38] and/or small nanoparticles down to the sub-10 nm size range, since they effectively enable the accumulation of reaction product over time to local concentrations high enough to enable their detection.

## Methods

### Instruments

A Nikon Eclipse LV150N upright microscope with a Nikon 50x ELWD dark-field objective was used to record the dark-field microscopy images and spectra. The light source was a Thorlabs Solis-3C LED lamp with an output power of 4 W. The scattered light from the nanochannels was spectrally resolved in an Andor grating spectrometer (SR-193I-A-SL) with a 150 l/mm grating and subsequently recorded with a Andor Newton (DU920P-BEX2-DD) camera attached to the spectrometer. The center wavelength of the spectrometer was set to 600 nm and the exposure time for each frame was 2 s. Using the multitrack-feature of the camera, the image was divided into 8 tracks, 2 for each reference side and 6 for the center nanochannel, with 3 on either side of the particle. Each track (21 px corresponding to 15 μm of nanochannel) was binned to give one spectrum per track. The constant flow of nitrogen in the microchannel was realized with a Fluigent MFCS-EX pressure controller, set to pressures 1500 mbar. The injection of water and dye solution was realized with syringes attached via tubes to the respective inlet reservoirs. ASP spectra of the dyes were recorded on a Varian Cary 500 spectrophotometer. The SEM images of the particles in the nanochannels were recorded on a Zeiss Supra 55VP scanning electron microscope. The pH values of the solutions were determined with a Hanna Instruments HI 2211 pH/ORP meter.

*Preparation of brilliant blue and fluorescein solutions*

The dyes used in this publication (Brilliant Blue and Fluorescein) were bought as their sodium salts from Merck as solid material and diluted into stock solutions of 50 mM with ultrapure water (Milli-Q IQ 7000 water purification, Merck). For each measurement, a fresh solution of $NaBH_4$ at 260 mM was prepared by mixing 10 mg of the substance with 1 ml of water. The dye stock solution and the reducing agent solution were then mixed in a 50/50 ration immediately before injecting them in the reservoirs of the chip holder. Injection into the fluidic chip holder was done with syringes and blunt needles (Braun).

*Fluidic chip fabrication*

The chips with the micro and nanofluidic systems used in the experiments were fabricated in the clean room facilities of MC2 at Chalmers in Gothenburg. Four of the octagonal fluidic chips could be produced from a 4-inch silicon wafer. This wafer was prepared with a thermal oxide layer into which the fluidic structures were later etched. The fabrication procedure of similar fluidic chips is described in more detail in our earlier work by Levin et al[37]. Short summary: The 4-inch (100) silicon wafers were cleaned with Standard Clean 1, followed by a 2% HF dip and Standard Clean 2. The thermal oxide layer was grown at 1050 °C in wet atmosphere until a thickness of the oxide of 250 nm was reached. The nanochannels were first patterned in a resist layer with electron beam lithography and then etched into the oxide via fluorine-based reactive ion etching (RIE). For the microchannels, photoresist was exposed with direct laser lithography and then etched with the same method. The inlet holes were etched through the wafer with deep reactive ion etching. The metal nanoparticles were placed in the nanochannel by patterning of a PMMA lift-off layer with electron beam lithography which was then developed to create a mask. The masked wafer was placed in a metal evaporator to create a layer of 20 nm gold, with every deposition except the particles being removed during the subsequent lift-off process. Finally, the substrate and a 175 µm thick Borofloat 33 glass wafer were treated with Standard Clean 1, with the glass wafer being the lid for the fluidic system. Both wafers were then treated with $O_2$ plasma (1 min, 50 W RF power, 250 mTorr) to facilitate the pre-bonding of the glass cover lid to the wafer with the fluidics. The final fusion bonding was carried out at 550 °C for 5h. After bonding the 4-inch fluidic wafer was cut into four octagonal chips to be used in the chip holder.

## Supporting Information

This material is available free of charge via the Internet at http://pubs.acs.org.

Incident light spectrum, RSID spectra of $NaBH_4$ solution, RSID and molar extinction coefficient spectra for BB-dye and fluorescein solutions; micro and nanofluidic operation scheme; batch reactor experiment with empty nanochannel without Au nanoparticle; ridge plot of NSS inverted RISD-spectra for section A of the nanochannel analyzed in Figure 5 in the main text; mass spectrum of the N2 gas used in the batch reactor experiments. (PDF)

The underlying data for this publication is available at Zenodo, DOI:10.5281/zenodo.11505293.

## Corresponding Author

Christoph Langhammer − Department of Physics, Chalmers University of Technology, SE-412 96 Gothenburg, Sweden; orcid.org/0000-0003-2180-1379; Email: clangham@chalmers.se

**Authors**

Björn Altenburger − Department of Physics, Chalmers University of Technology, SE-412 96 Gothenburg, Sweden; orcid.org/0009-0003-0600-4635

Joachim Fritzsche − Department of Physics, Chalmers University of Technology, SE-412 96 Gothenburg, Sweden; orcid.org/0000-0001-8660-2624

**Author Contributions**

The manuscript was written through contributions of all authors. All authors have given approval to the final version of the manuscript.

**Acknowledgements**

This research has received funding from the Swedish Research Council (VR) Consolidator Grant project 2018-00329 and the European Research Council (ERC) under the European Union's Horizon Europe research and innovation program (101043480/NACAREI). Part of this work was carried out at the Chalmers MC2 cleanroom facility and at the Chalmers Materials Analysis Laboratory (CMAL). We also acknowledge fruitful discussions with Dr. Jordi Piella, Puvanesvari Teluchina-Appadu, Prof. Henrik Sundén and Prof. Anders Hellman.

# Supplementary Material for

## Femtoliter Batch Reactors for Nanofluidic Scattering Spectroscopy Analysis of Catalytic Reactions on Single Nanoparticles

*Björn Altenburger[1], Joachim Fritzsche[1] and Christoph Langhammer[1\*]*

[1]Department of Physics, Chalmers University of Technology; SE-412 96 Gothenburg, Sweden

*Corresponding author: clangham@chalmers.se

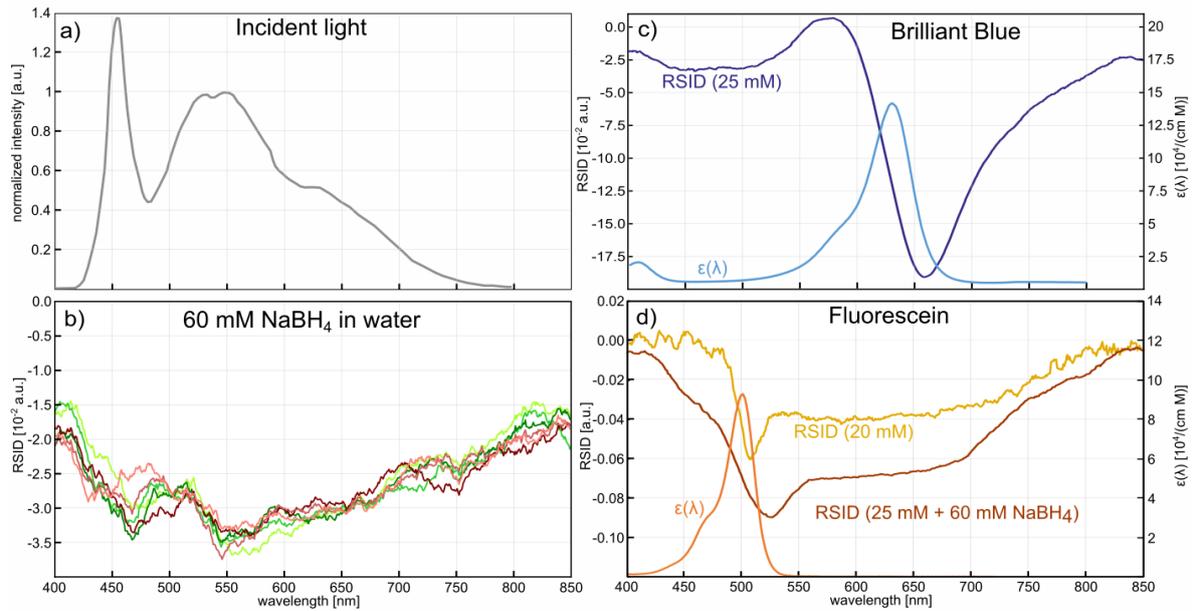

**Figure S1. Collection of spectra relevant for this work.** a) Emission intensity spectrum of the light source (Thorlabs Solis 3C) used for the NSS measurements. b) RSID spectrum of a 60 mM NaBH4 solution measured in a nanochannel. The color code depicts the six nanochannel sections used for NSS readout and is the same as introduced in **Figure 3** in the main text. c) RSID spectrum of a 25 mM aqueous Brilliant Blue solution together with the molar extinction coefficient spectrum, $\varepsilon(\lambda)$, calculated from the RSID spectrum using the formalism we have introduced earlier[1]. We note that the main peak of the $\varepsilon(\lambda)$ spectrum corresponds to the negative slope of the RSID spectrum. d) RSID spectra of a 20 mM and 25 mM Fluorescein solution, where the latter has been mixed with 60 mM NaBH$_4$ to increase the pH value. We note the higher amplitude of the negative RSID peak in alkaline environment compared to the one in just MilliQ water, as well as the distinct shift to longer wavelengths at the higher pH. Also plotted is a Fluorescein $\varepsilon(\lambda)$ spectrum measured using a spectrophotometer. We again note that the $\varepsilon(\lambda)$ peak maximum is localized at the negative slope of the negative RSID peak of the aqueous fluorescein solution, as expected.

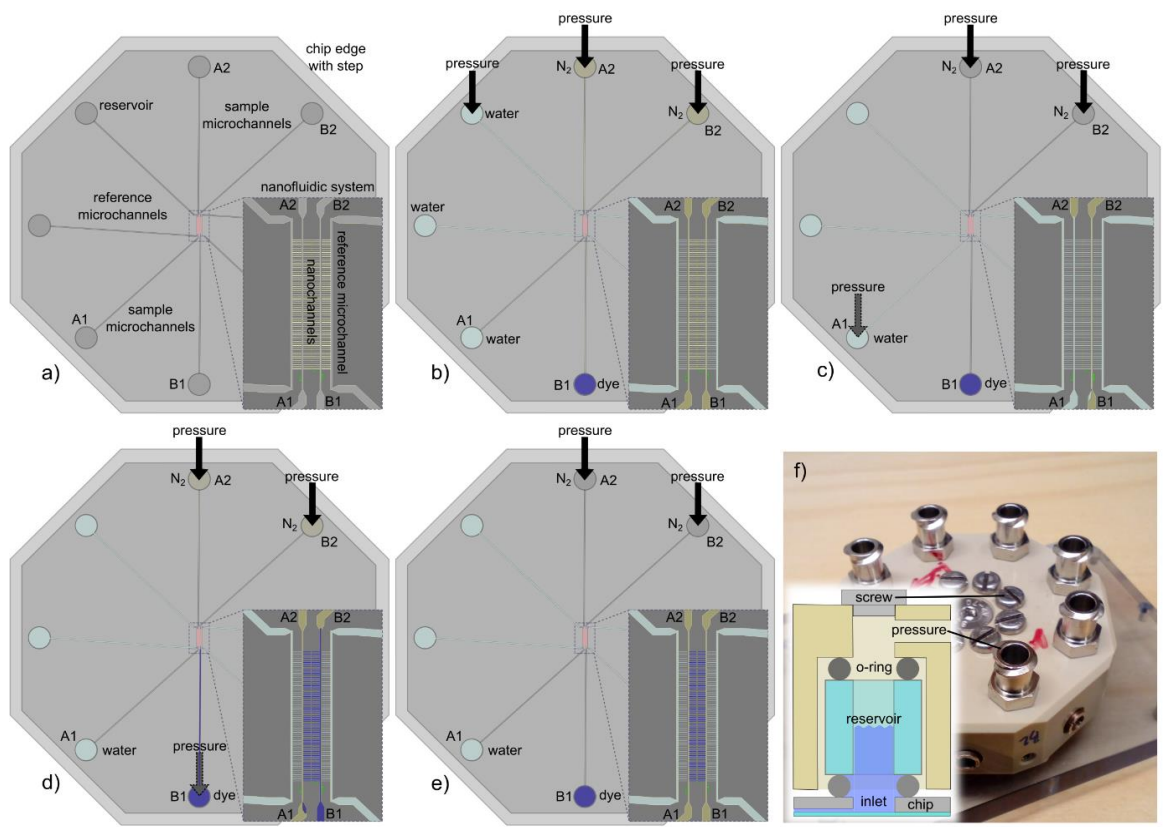

**Figure S2. Micro and nanofluidic operation scheme.** a) The fluidic chip is a silicon-based platform that comprises micro- and nanochannels, where the former connect to inlet reservoirs on one end and to the nanochannel system in the center (inset) on the other end. After fabrication, the entire fluidic system is filled with air. b) When implementing the batch reactor according to the steps summarized in **Figure 4a-d** in the main text, to execute the first step, the reservoirs of the reference microchannels are filled with water and pressurized with $N_2$ gas to create a flow of water through the reference microchannels to fill the reference nanochannels with water (see inset). At the same time, inlet A1 is filled with water and inlet B1 is filled with the BB-dye solution. To keep the liquids in inlets A1 and B1 from entering the sample nanochannel system too early, $N_2$ gas is flushed in through inlets A2 and B2 by applying a pressure of 1500 mbar. c) For the second step, measuring the intrinsic difference spectra with NSS, the central nanochannels are flushed with water (**cf. Figure 4b**). To do so, pressure is applied to the inlet reservoir A1 with a syringe, thereby counteracting the still applied $N_2$ pressure in the microchannel until the water reaches the nanochannels. The overpressure on the water side and the capillary forces causes the water to flush through the nanochannels (see inset) and then be carried away by the $N_2$ gas flow on the exit side. d) To flush the dye solution into the nanochannels in the third step, (**cf. Figure 4c**), the pressure on A1 is released and instead applied to B1. It is here of importance to limit the extension of the dye solution to the thinner sections of the microchannel (as shown in the inset), as an extension into the microchannel towards B2 could cause solution pockets that disturb the concentration in the nanochannels later. It is also recommended that the nanochannels are flushed thoroughly with the dye solution before the next step, as the water that has been flushed through to the dye side and has remained in the microchannel toward B1 changes the concentration of the first part of the dye solution when it is flushed in. e) To close off the nanochannels and establish the batch reactor condition in the fourth step (**cf. Figure 4d**), the pressure on B1 is released. The $N_2$ pressure on B2 pushes the liquid out of the thinner sections of the microchannel while the nanochannels remain filled with the dye solution (inset). If it is of interest to repeat the procedure and fil a batch reactor once again, it is necessary to exchange the liquids in reservoirs A1 and B1 since they are contaminated. f) The octagonal chips are clamped in a holder that provides a short glass tube for each inlet to function as reservoirs. These tubes are tightened to the chip via O-rings (see inset). Liquid can be inserted via a screw over the reservoirs while pressure is applied via side connections and Luer-Lock couplings, using a Fluidgent device or a syringe.

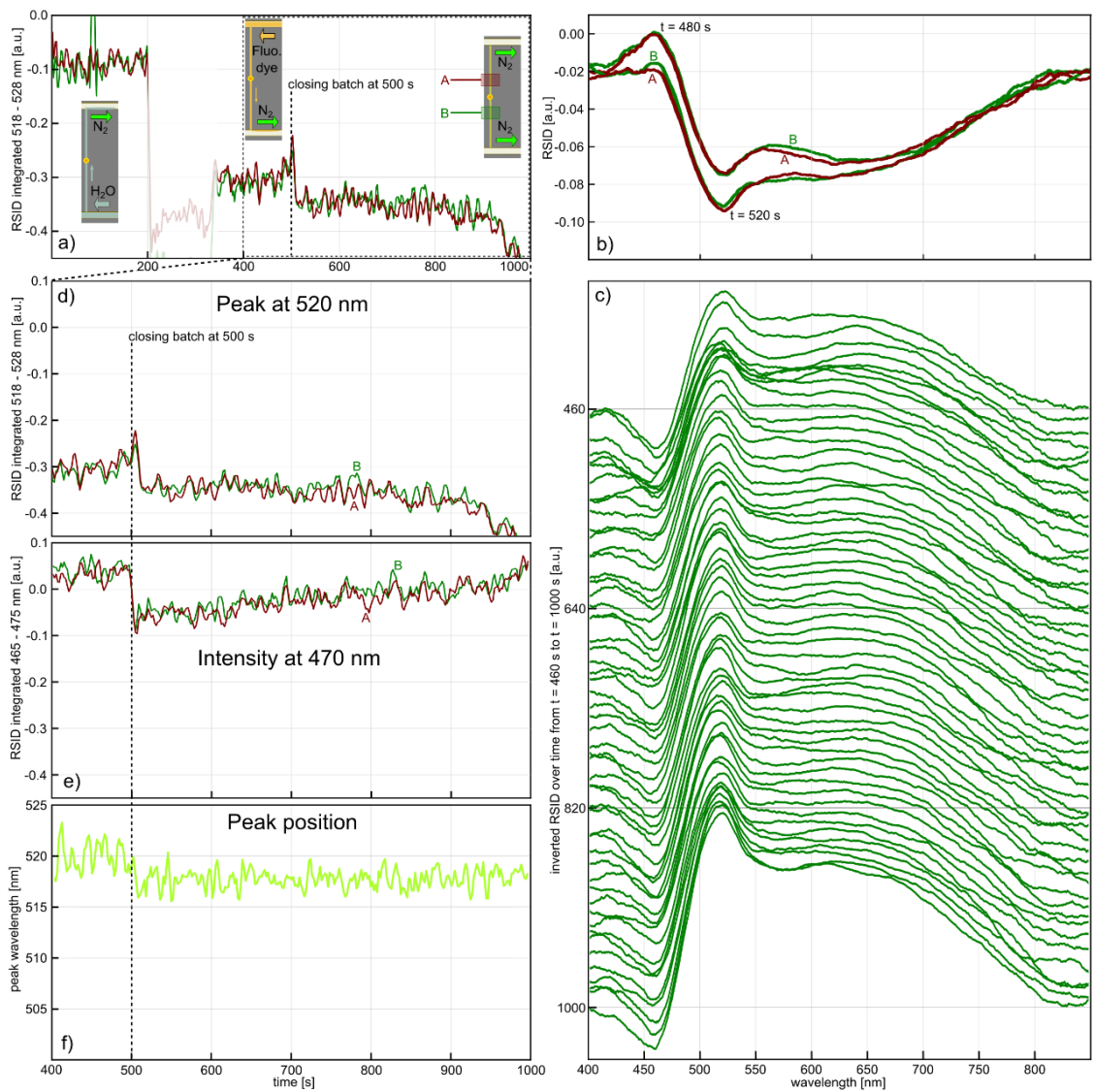

**Figure S3. Batch reactor experiment with empty nanochannel without Au nanoparticle.** a) Time trace of the integrated RSID value between 518 – 528 nm for the same two nanochannel sections as depicted in **Figure 5** in the main text. The same three subsequent steps in the batch reactor operation are indicated along the time trace: i) flushing of water (0 – 200 s) flushing of the 25 mM fluorescein and 120 mM NaBH₄ solution (350 – 500 s), closed batch reactor (500 – 1000 s). The shaded area marks the time where monitoring of the sample through the microscope eyepiece was necessary. b) RSID spectra for the two sections of the nanochannel at a time before (480 s) and after (540 s) the batch reactor was closed at 500 s. c) Ridge plot of the inverted RSID spectra taken at 10 s intervals from t = 460 s to 1000 s. In contrast to **Figure 5c** in the main text, where the Au nanoparticle is present in the nanochannel, no significant change of the RSID spectrum occurs and no new short-wavelength peak emerges. d) Zoom-in of the time trace depicted in a) that follows the integrated amplitude of the main RSID peak, that corresponds to the fluorescein absorption band in alkaline environment. We attribute the slight increase of negative peak amplitude towards more negative values in the final stages of the experiment to the onset of channel drying, as discussed in **Figure 4** in the main text. e) Same as d) but when integrating the RISD between 465 nm – 475 nm, i.e., in the same range where the second peak emerged in the experiment with Au nanoparticle. Clearly, it is absent in this control experiment. The step seen at 500 s (directly upon closing of the batch reactor) in the time traces in both d) and e) are due to slight defocusing of the stage. f) Spectral position of the main negative RSID-peak at 520 nm that corresponds to the fluorescine absorption band in alkaline conditions over time. We note that it, in contrast to the corresponding analysis in the presence of the Au nanoparticle depicted in **Figure 5f** in the main text, essentially stays constant.

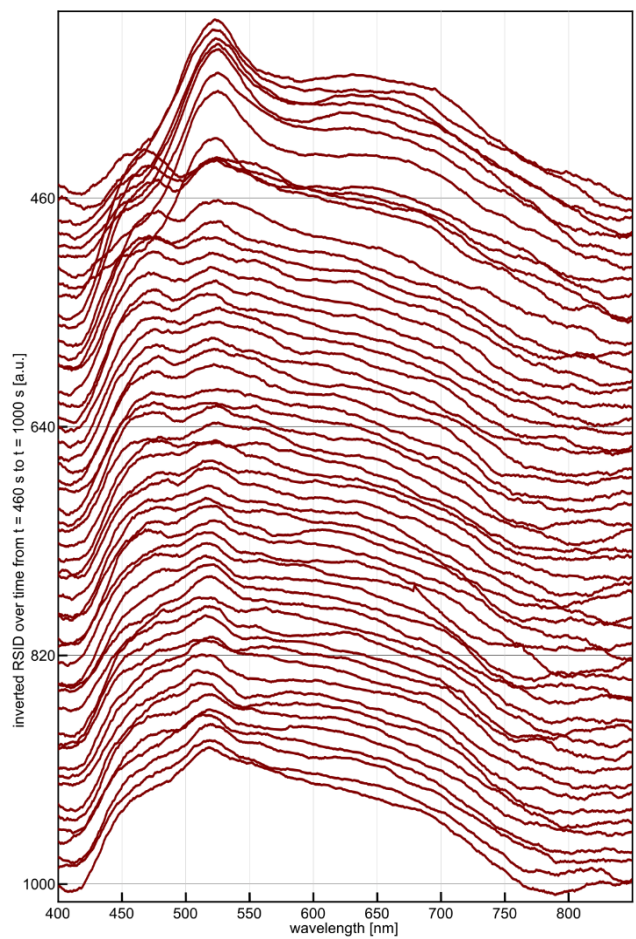

**Figure S4. Ridge plot of NSS inverted RISD-spectra for section A of the nanochannel analyzed in Figure 5 in the main text.** The spectra plotted start at t = 460 s (reactant solution is flowing through the nanochannel) and are recorded up to 1000 s (batch reactor has been closed and working for 480 s).

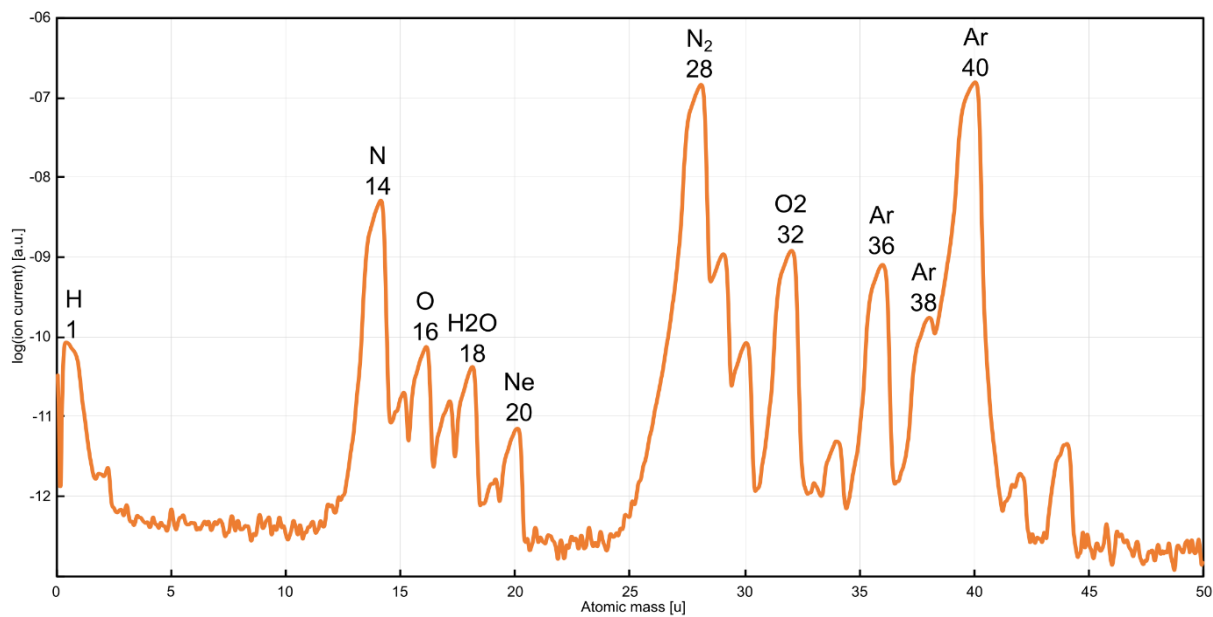

**Figure S5. Mass spectrum of the N₂ gas used in the batch reactor experiments.** The spectrum was taken with a Pfeiffer OmniStar GSD 320 gas mass spectrometer. Apart from the monoatomic (14 u) and diatomic (28) nitrogen fragments, there are various other gases present in the steam. Most prominent among them is Argon and its isotopes (36 u, 38 u, 40 u), but there are also traces of hydrogen (1 u) $H_2O$ (18 u) and Neon (20 u). In addition, and most importantly here, we also measure distinct signals that correspond to the monoatomic (16 u) and diatomic (32 u) fragments of oxygen, which corroborate the presence of sizable amounts of $O_2$ in the $N_2$ gas stream.